# Local Manipulation of Skyrmion Lattice in $Fe_3GaTe_2$ at Room Temperature

Shuaizhao Jin [a], Zhan Wang [b], Shouzhe Dong [c], Yiting Wang [d,e], Kun Han [d,e], Guangcheng Wang [f], Zunyi Deng [a], Xingan Jiang [a,g], Ying Zhang [b], Houbing Huang [c], Jiawang Hong [a], Xiaolei Wang [f,*], Tianlong Xia[d,e,*], Sang-Wook Cheong [h], Xueyun Wang [a.*]

[a] School of Aerospace Engineering, Beijing Institute of Technology, Beijing 100081, China

[b] Beijing National State Key Laboratory of Magnetism, Institute of Physics, Chinese Academy of Sciences; Beijing 100190, China

[c] Advanced Research Institute of Multidisciplinary Science, and School of Materials Science and Engineering, Beijing Institute of Technology, Beijing 100081, China

[d] Beijing Key Laboratory of Opto-Electronic Functional Materials & Micro-Nano Devices, Department of Physics, Renmin University of China, Beijing 100872, China

[e] Key Laboratory of Quantum State Construction and Manipulation (Ministry of Education), Renmin University of China, Beijing 100872, China

[f] Department of Physics and Optoelectronic Engineering, Faculty of Science, Beijing University of Technology, Beijing 100124, China

[g] Institute of Micro/Nano Materials and Devices, Ningbo University of Technology, Ningbo City 315211, China

[h] Rutgers Center for Emergent Materials and Department of Physics and Astronomy, Rutgers University, Piscataway, New Jersey 08854, USA

[*]Correspondence to E-mail: xiaoleiwang@bjut.edu.cn; tlxia@ruc.edu.cn; xueyun@bit.edu.cn.

## Abstract

Motivated by advances in spintronic devices, an extensive exploration is underway to uncover materials that host topologically protected spin textures, exemplified by skyrmions. One critical challenge involved in the potential application of skyrmions in van der Waals (vdW) materials is the attainment and manipulation of skyrmions at room temperature. In this study, we report the creation of intrinsic skyrmion state in van der Waals ferromagnet $Fe_3GaTe_2$. By employing variable temperature magnetic force microscopy, the skyrmion lattice can be locally manipulated on $Fe_3GaTe_2$ flake. The ordering of skyrmion state is further analyzed. Our result suggest $Fe_3GaTe_2$ emerges as a highly promising contender for the realization of skyrmion-based layered spintronic memory devices.

## 1. Introduction

Magnetic skyrmion, a particle-like non-trivial spin structure, have attracted extensive attention in spintronics field due to their peculiar physical properties [1,2]. This spin texture with nano-size, topology-preserving stability and low driving current density exhibits great potential for applications in novel magnetic memory devices [3–5]. Magnetic skyrmion was initially observed in non-centrosymmetric layer bulk. The bulk Dzyaloshinskii-Moriya interaction (DMI) activated by symmetry breaking plays a crucial role in stabilizing the skyrmion structure and imparts the magnetic structure with stable chiral property [6]. Subsequent studies demonstrate that skyrmions can be generated in magnetic multilayer structures embedded in non-magnetic heavy metal layers [7–11], where the DMI is mainly triggered by interface asymmetry. Apart from non-centrosymmetric chiral magnets and multilayers, skyrmions can also be formed in centrosymmetric materials with strong uniaxial magnetic anisotropy (UMA), such as $La_{1-x}Sr_xMnO_3$ ($x$ = 0.175) [12], MnNiGa [13], and frustrated magnet $Fe_3Sn_2$ [14], in which the competition between dipole-dipole interaction and magnetocrystalline anisotropy dominate.

The discovery of long-range magnetic ordering in van der Waals (vdW) magnets marks a significant advancement in the exploration of novel physical properties and topological magnetic textures [15–19]. In the study of vdW $Fe_3GeTe_2$ featuring UMA, the revelation of skyrmions has spurred substantial research endeavors focusing on topological spin textures [20]. Recently, the emergence of composite skyrmions was only achieved in $Fe_{3-x}GeTe_2$ below room temperature [21]. The dipole-dipole interaction and the DMI effects presented at oxidized interface give rise to the Néel- and Bloch-type skyrmions in $Fe_3GeTe_2$ samples with a centrosymmetric structure [22,23]. Notably, the transformation of Néel- and Bloch-type skyrmions can be manipulated by varying the Fe content and thickness of $Fe_{3-\delta}GeTe_2$ [24]. In addition, the heterostructures of $WTe_2/Fe_3GeTe_2$, oxidized-$Fe_3GeTe_2/Fe_3GeTe_2$ and $Cr_2Ge_2Te_6/Fe_3GeTe_2$, under the action of interface DMI, have been evidenced to produce stable Néel-type skyrmions, highlight the crucial of the interfacial DMI in generating magnetic skyrmions [25–27].

The recent discovery of vdW ferromagnetic materials at room temperature provides a research platform for room temperature skyrmion. For example, Co doping can increase the curie temperature ($T_c$) of FGT to 363 K [28], after which room-temperature skyrmion lattices were observed in $(Fe_{0.5}Co_{0.5})_5GeTe_2$ [29]. A recent study demonstrated that layered $Fe_3GaTe_2$ vdW single crystal can maintain intrinsic ferromagnetism and variable perpendicular magnetic anisotropy (PMA, a special type of UMA) above room temperature [30]. Skyrmion-like magnetic bubbles have been identified at room temperature in $Au/Fe_3GaTe_2$ heterostructures [31], but their magnetic structure has not been determined by characterization. Furthermore, the existence of skyrmions in $Fe_3GaTe_{2-x}$ single crystals from 2 K to 300 K is predicted by the topological Hall effect [32]. However, in $Fe_3GaTe_2$, there is currently no direct real-space imaging evidence confirming the existence of intrinsic skyrmions state at room temperature. Therefore, a more comprehensive investigation is necessary to explore the capacity, conditions, and mechanisms involved in generating room temperature skyrmions for $Fe_3GaTe_2$.

In this article, employing methods of real-space imaging, we have successfully created intrinsic skyrmions on $Fe_3GaTe_2$ at room temperature through three approaches: magnetic field inducting, field cooling (FC), and magnetic force microscopy (MFM) scanning. Notably, in combination with Lorentz transmission electron microscopy (L-TEM) and phase-field simulations, we identified the magnetic structure and generation mechanism of Néel-type skyrmions. Additionally, we propose a method for precise local manipulation of skyrmions using an MFM tip. This approach enables reversible creation and erasure of skyrmions at arbitrary positions on the sample, presenting an innovative write-erase technique for skyrmions used as magnetic storage units.

## 2. Experimental

### 2.1 Sample Preparation

In this work, $Fe_3GaTe_2$ single crystal was grown using the self-flux method. High-purity Fe powder (99.99%) Ga lumps (99.99%) and Te powder (99.999%) were mixed at atomic ratios of 1:1:2 in a sealed quartz tubes with high vacuum, in which the Ga and Te are also acting as the flux. The quartz tubes are placed in a muffle furnace and heated to 1273 K in 1 h. After holding for 24 h, the temperature is rapidly lowered to 1153 K in 1 h. Then, the temperature is slowly lowered to 1053 K over a period of 120 h. Finally, single crystals of $Fe_3GaTe_2$ can be obtained by centrifugation.

2.2 Characterization techniques

The morphology, crystal structure, elemental distribution and ratios of $Fe_3GaTe_2$ were investigated using optical microscopy (OM, Axio Imager A2m), powder X-ray diffraction (XRD, Rigaku Ultima IV) with Cu Kα radiation (wavelength = 0.15406 nm) and Field Emission Scanning Electron Microscope (FE-SEM, TESCAN MIRA LMS) equipped with energy dispersive X-ray spectroscopy (EDS), respectively. These measurements were performed at room temperature. Magnetic measurements of $Fe_3GaTe_2$ crystals were carried out employing a Quantum Design magnetic property measurement system (QD MPMS-3). The magnetic domains were studied by using a JEOL-dedicated L-TEM (JEOL 2100F) and an atomic force microscope (Asylum Research MFP-3D Classic). Comprehensive details on magnetic domain characterization, device fabrication, and Hall measurement can be found in the Supplemental files.

## 3. Results

3.1. Crystal structure and magnetic properties

High quality $Fe_3GaTe_2$ single crystal with typical lateral sizes of 3 × 2 $mm^2$ (inset in Fig. 1d) was grown using the self-flux method. The crystal structure of the $Fe_3GaTe_2$ is shown in Fig. 1a. Similar to $Fe_3GeTe_2$, Fe and Ga atoms are encapsulated by two layers of Te atoms in the $Fe_3GaTe_2$ layered structure. Two unique Fe Wyckoff positions are occupied by $Fe_I$ (in the top and bottom planes) and $Fe_{II}$ (in the central plane) atoms, where $Fe_I$-$Fe_I$ layers form triangular lattices while $Fe_{II}$-Ga layers form hexagonal network. The 2D vdW $Fe_3GaTe_2$ crystallizes in hexagonal structure with a centrosymmetric space group of $P6_3/mmc$. Fig. 1b shows the EDS mapping images of $Fe_3GaTe_2$. The elements, with strongly colored Fe, Ga, and Te, are distributed uniformly. Further, EDS spectra in Fig. 1c exhibits the elemental ratios of Fe, Ga, and Te as 2.93:1.07:2.00. As shown in Fig. 1d, the XRD peaks of $Fe_3GaTe_2$ agree highly with previously reported [30], indicating a high quality of the single crystals grown in this study.

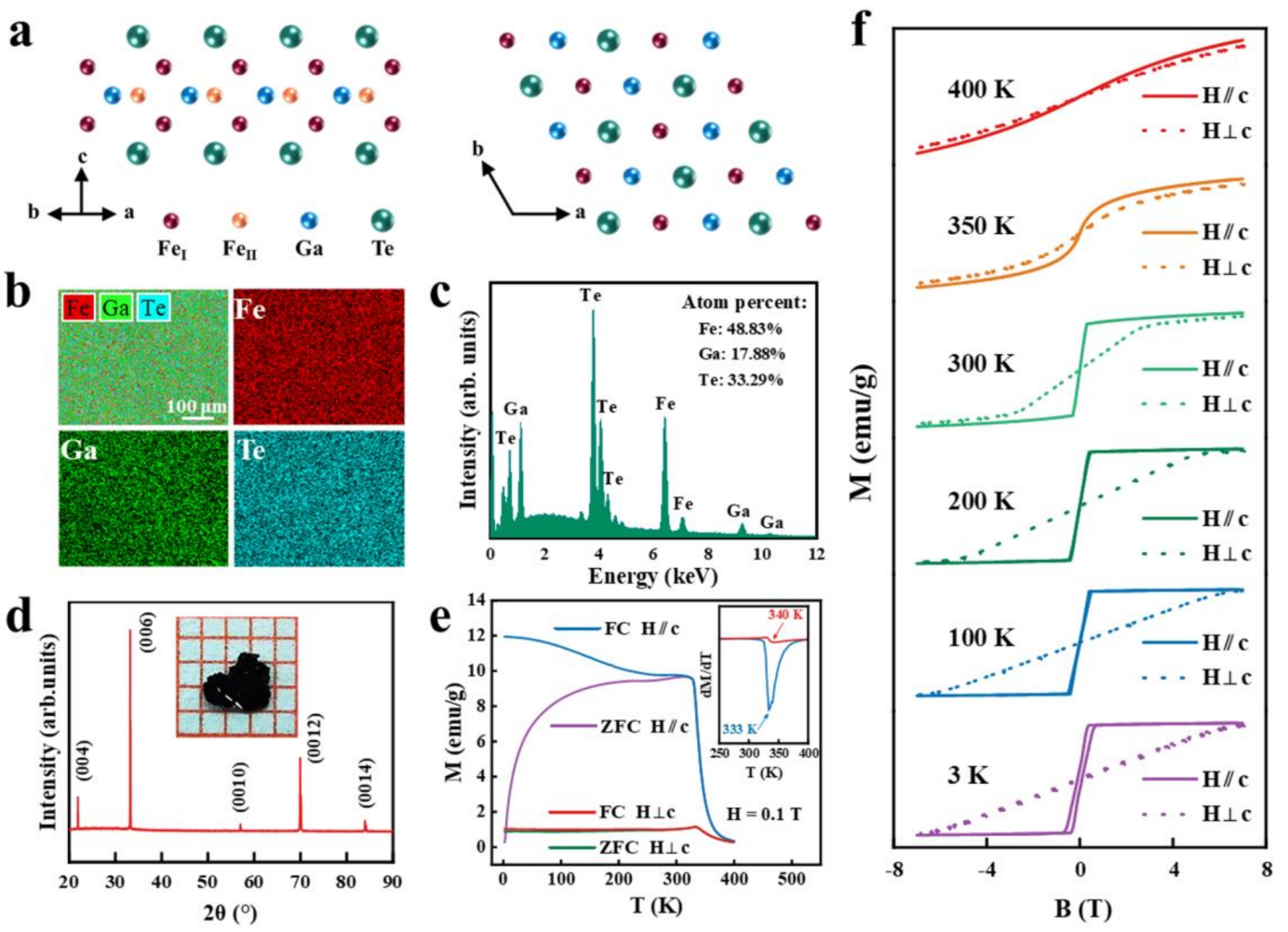

**Fig. 1.** Crystal characterization and magnetization measurements of $Fe_3GaTe_2$ single crystals. (a) Top view (left) and side view (right) of $Fe_3GaTe_2$ crystal structure. EDS (b) and elements mapping (c) of a $Fe_3GaTe_2$ specimen. (d) Experimental XRD spectra of $Fe_3GaTe_2$. The inset shows a single crystal image, the grid size is 1 $mm^2$. (e) Temperature dependence of magnetization for ZFC and FC curves at 0.1 T perpendicular and parallel to the *c*-axis. The inset shows the derivative of the FC $M(T)$ curves. (f) Magnetization of $Fe_3GaTe_2$ single crystal perpendicular (solid) and parallel (dashed) to the *c*-axis at different temperatures.

In order to investigate the intrinsic magnetic properties, $Fe_3GaTe_2$ single crystal was characterized using vibrating sample magnetometer (VSM). As shown in Fig. 1e, temperature-dependent magnetization curves, at external fields $H$ = 0.1 T perpendicular ($H \perp$ c) and parallel ($H /\!/$ c) to the *c*-axis, were obtained under zero-field cooling (ZFC) and FC. The derivative of *M-T* curves (d*M*/d*T*) for FC ware plotted in the inset to derive $T_c$. The minimum values of d*M*/d*T* for the FC *M-T* curves with $H \perp$ c and $H /\!/$ c are 340 K and 333 K, respectively, indicating the $T_c$ of the $Fe_3GaTe_2$ is within the 333-340 K range. The magnetization of *H*//c is significantly greater than that of $H \perp$ c, showing an easy-axis magnetic anisotropy for the flakes [30]. The FC and ZFC curves of *H*//c separate at temperatures below $T_c$, which is similar to other typically frustrated magnets [33–35]. In addition, the hysteresis loops at different temperatures with $H \perp$(dashed lines) cand *H*//c (solid lines) are shown in Fig. 1f. Large PMA of $Fe_3GaTe_2$ is evidenced by the difference in the M-H curves of $H \perp$ c and *H*//c. This difference diminishes as the temperature increases, indicating a gradual weakening of PMA. At room-temperature of 300 K, $Fe_3GaTe_2$ crystal exhibits the PMA energy density ($K_u$) and saturation magnetic moment around $4.15\times10^5$ J/m$^3$ (detailed calculations are provided in the supporting information) and 33.26 emu/g (0.98 μB/Fe) respectively, which is close to previous reports [30].

### 3.2 Real-space characterization of skyrmions

L-TEM was employed to real-space characterize the magnetic domain structure of $Fe_3GaTe_2$. Fig. 2a-d show the field-dependent evolution of the magnetic domains observed at room temperature (295 K). In these measurements, $Fe_3GaTe_2$ flake with a thickness approximately 175 nm was horizontally tilted by $\alpha = -8°$ and the defocus value was −2 mm. Labyrinthine domain structure can be observed in Fig. 2a without an external magnetic field. As shown in Supplementary Fig. S1, these labyrinthine domains are only visible at $\alpha \neq 0$, indicating a Néel-type nature of the domain walls. Within a magnetic field range of 0 ~ 1998 Oe, the labyrinthine domains gradually transform into skyrmion-like bubbles as the magnetic field increases. The skyrmion-like bubbles start nucleating at 1338 Oe. (Fig. 2b). In Fig. 2c, most of the labyrinthine domains are transformed in stripe domains, worm-like magnetic domains and skyrmion-like bubbles at a field of 1574 Oe. As the magnetic field further increases, a vanishing contrast in Fig. 2d suggests that the sample is magnetized to a ferromagnetic state.

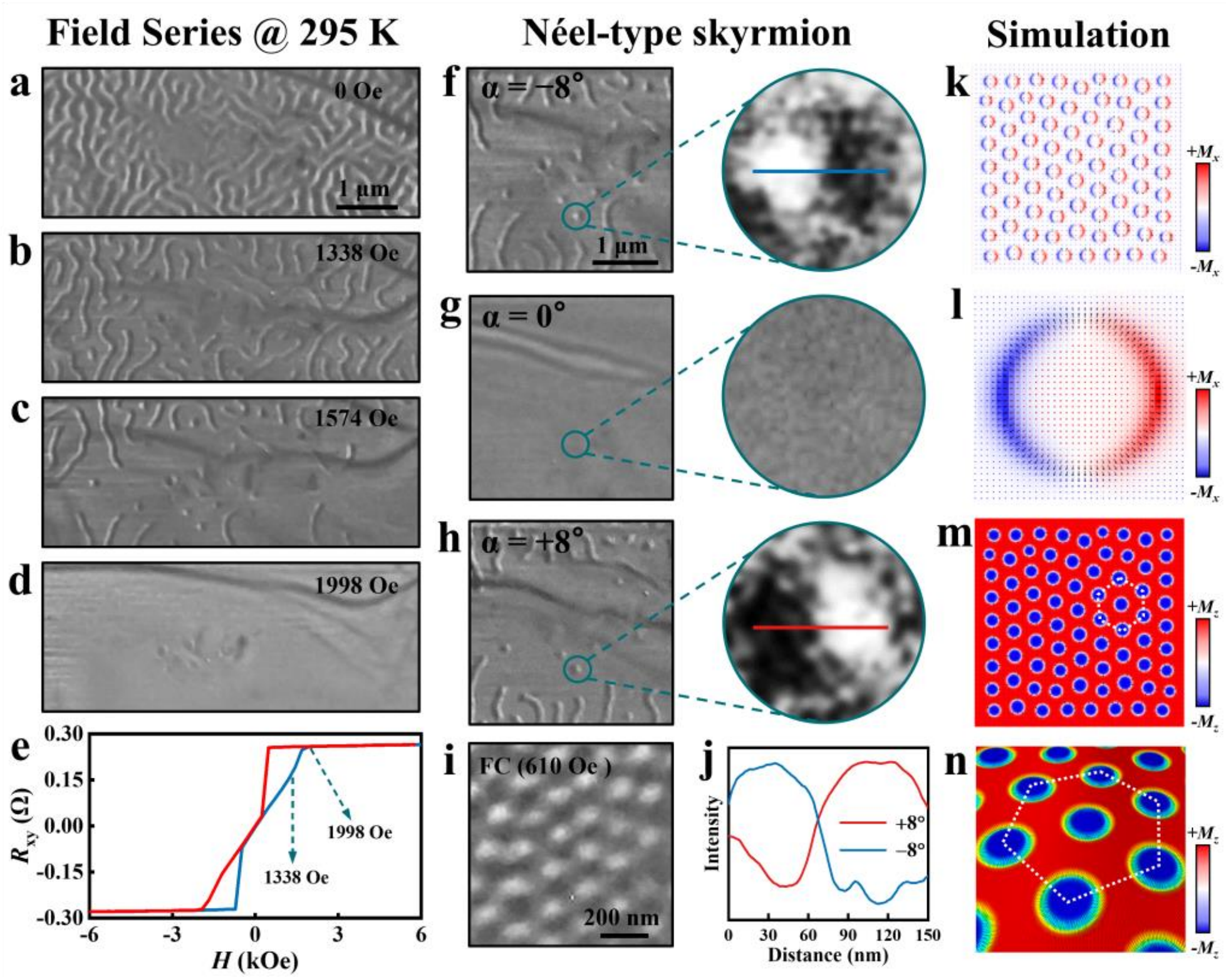

**Fig. 2.** L-TEM and phase field simulation of Magnetic domains. (a-d) Evolution of magnetic domains in various magnetic fields. **(**e) Hall resistance $R_{xy}$ as a function of applied out-of-plane magnetic field, The blue (red) line represents the direction of increasing (decreasing) magnetic field. **(**f-h) L-TEM images and enlarged skyrmion views with different sample tilt, α, at 1574 Oe field. (i) Skyrmion lattice observed after cooling at 610 Oe magnetic field. (j) Line profiles of the blue and red lines labelled in (f) and (h). (k-n) Simulation of skyrmion lattices at an applied magnetic field of 600 Oe and a DMI value of 0.4 $mJ/m^3$. The magnetization along the *x*-axis (k) and *z*-axis (m) is indicated by the red region ($+M_x$ or $+M_z$) and the blue region ($-M_x$ or $-M_z$), respectively. (l) Enlarged view of the magnetic configuration for a single Néel-type skyrmion in (k). (n) Enlarged schematic of skyrmion lattice magnetic configuration of (m).

The exfoliated $Fe_3GaTe_2$ flake was processed into devices, and an out-of-plane magnetic field was applied to obtain the Hall resistance ($R_{xy}$). The hysteresis characteristic of the Hall measurement indicates a typical topological property, as depicted in Fig. 2e. It is worth noting that there are two different slopes on the Hall curve before the magnetic field increases (blue line) from 0 Oe to saturation. The transition point in the Hall curves exhibiting distinct slopes aligns with a magnetic field approaching 1338 Oe, corroborating the emergence of the skyrmion-like bubble, as depicted in Fig. 2b. Furthermore, the Hall curve's saturation point approximates 1998 Oe, in agreement with the field at which the domains vanish in Fig. 2d. As a result, the initial slope of the Hall curve is associated with labyrinthine domains, while the subsequent slope corresponds to multidomain states. This observation confirms the topological texture, which is in accordance with previous research [26].

A tilt series of 8°, 0° and -8° taken from a local image in Fig. 2c is shown in Fig. 2f-h. No distinct contrast is evident at 0° tilt angle (with the electron beam perpendicular to the sample surface). However, upon tilting the sample by α = ±8°, bubbles with well-defined boundary becomes discernible. As visible in the enlarged image at the same location, the bubble exhibits a contrast with half-dark and half-bright, which is inverted as the tilt angle is reversed. These phenomena demonstrate that the magnetic bubbles produced in $Fe_3GaTe_2$ are Néel-type skyrmions [36]. Detailed intensity inversion information for Néel-type skyrmion at opposite tilt angles is shown in the line profiles of Fig. 2j. In the phase field simulation below, we show that the interface DMI in this system is critical to stabilize the Néel-type skyrmions. Note that magnetic clusters emerging above the $T_c$ act as the precursors of skyrmion [37,38]. During the FC process, a magnetic field applied above $T_c$ promotes the stabilization and organized growth of magnetic clusters characterized by short-range ordering [39]. This is a pivotal factor in nucleating high-density skyrmion lattice. Therefore, skyrmion lattice can be generated after FC (see Fig. 2i and Fig. S2). As shown in Supplementary Fig. S3, the skyrmion lattice was not detected at α = 0°, whereas can be observable at α = ±13°, suggesting that skyrmions generated after FC also exhibit Néel-type behavior. Moreover, as illustrated in Supplementary Fig. S4, the skyrmion density is approximately constant during sample cooling under the magnetic field of 300 and 600 Oe.

$Fe_3GaTe_2$ crystalizes in a centrosymmetric space group without symmetry breaking and DMI. However, the surface of metallic vdW material can be slightly oxidized during the exfoliation. The interfacial DMI induced by the oxide layer in $Fe_3GaTe$ has been identified as the key factor in generating the topological Hall effect [32]. Therefore, it is predicted that Néel-type skyrmions can be formed under the stabilization of DMI generated at the interface between oxide layer and $Fe_3GaTe_2$. In order to verify the above, phase field simulations were carried out under different conditions shown in Supplementary Fig. S5. In the absence of DMI, the magnetic structure exhibits Bloch-type labyrinthine domains and skyrmions at 0 and 600 Oe magnetic fields, respectively. At the magnetic field of 600 Oe, skyrmions gradually transform from Bloch-type to Néel-type after introducing DMI. As shown in Fig. 2k-m, under the effect of 0.4 mJ/m$^3$ DMI value, Néel-type skyrmions were obtained in the simulations. This skyrmion observed in Fig. 2n constitutes a lattice in the form of a sixfold coordination. Thus, the Néel-type skyrmion in $Fe_3GaTe_2$ exhibited at room temperature is ascribed to the competition between Heisenberg exchange interaction, interface DMI, and magnetocrystalline anisotropy.

### 3.3 Precise creation and erasure of skyrmion lattice

According to the L-TEM results and previous studies [40–42], skyrmion lattice can be generated by FC. FC can also be achieved in MFM with the assistance of heating device, in which a local magnetic field is provided by the magnetic scanning tip (as depicted in Fig. 3a). The cobalt alloy coated tip exhibits a magnetic moment of approximately $10^{-13}$ emu and a coercivity of roughly 300 Oe. Fig. 3b illustrates the creation of skyrmion lattice. According previous studies, the free energy of the skyrmion lattice is higher than labyrinthine domain phase, resulting in an energy barrier between the two phases [29,43]. The labyrinthine domain is stable between 30~50 °C. As the temperature rises to 60 °C, the labyrinthine domains start to split into worm-like and bubble-like domains. Driven by a combination of thermal activation effect and the Zeeman energy gain of the skyrmion lattice phase, a transformation from labyrinthine domains to skyrmion emerges. Labyrinthine domains completely transform into metastable skyrmion lattice at 65 °C. Further increase in temperature leads to skyrmion lattice disappearance at 75 °C. A narrow temperature window exists for the skyrmion phase, which is consistent with previous studies [23,29]. At 80 °C, no magnetic domain is observed owing to the paramagnetic phase. In contrast to the field heating, the magnetic domain and skyrmion during FC appear at 75 °C and 70 °C, respectively. As further decreases from 65 °C, the skyrmion tends to be stable and the lattice gradually densifies. Therefore, a clear skyrmion lattice can be observed at 30 °C. Notably, skyrmion lattice produced by FC does not reconvert to labyrinthine domains as the temperature falls back to room temperature, which is attributed to an increase in energy barriers during the FC process. The skyrmion lattice remains stable after multiple scanning, as shown in Supplementary Fig. S6.

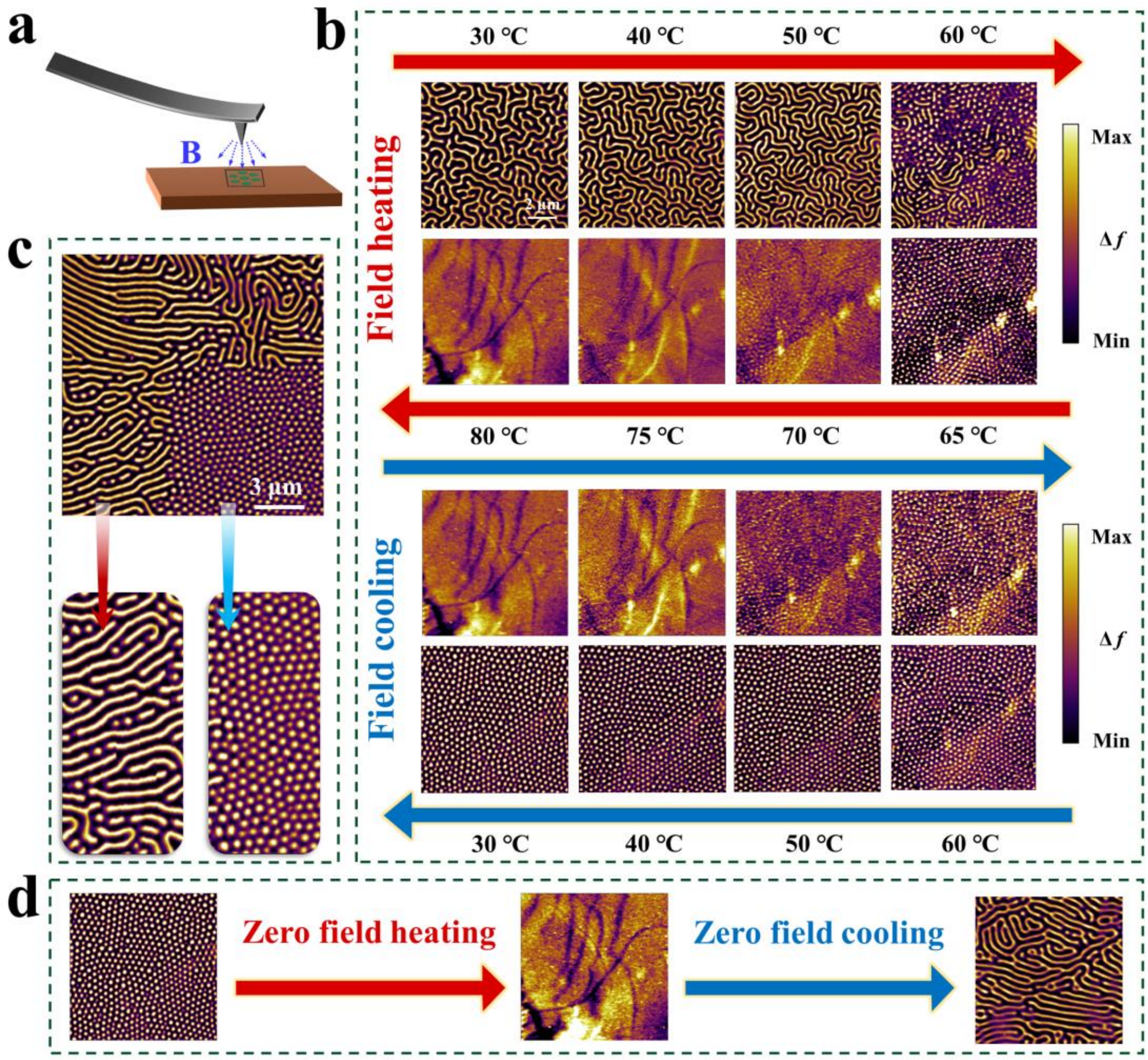

**Fig. 3.** Creation and erasure of skyrmion lattice by utilizing MFM scanning tip. (a) Schematic diagram of the magnetic field provided by the MFM tip. (b) Creation of skyrmion lattice under (MFM tip's) field heating and cooling. (c) skyrmion lattice created by MFM local scans with field heating and cooling. (d) Erasure of skyrmion lattice under zero field (after tip removal) heating and cooling.

To confirm the importance of the magnetic field provided by the scanning tip for generating skyrmion lattices, a fixed region was selected for MFM scanning during heating and cooling, with peripheral regions served as a comparison. As shown in Fig. 3c, a skyrmion lattice is generated in the scanned region under the action of MFM tip, whereas the peripheral control region still exhibits labyrinthine domains. It suggests that the MFM tip is necessary for inducing skyrmion generation. Precise generation of skyrmion lattice can be achieved by sweeping a fixed region while heating and cooling samples.

The removal of skyrmion lattice can also be achieved as shown in Fig. 3d. The MFM tip's lift height is elevated to a level where its magnetic field minimally influences the flakes. Subsequently, the sample undergoes heating and cooling processes, realizing both zero-field heating and zero-field cooling conditions. As the sample is heated to 80 °C at zero-field, the skyrmion lattice transform to a paramagnetic phase. Due to the absence of magnetic field provided by MFM tip, magnetic domains returned to labyrinthine domains after cooling the sample to room temperature at zero-field.

### 3.4 Ordering of skyrmions

In the skyrmion lattice generated by field heating and cooling (Fig. 3b), the skyrmion arrangements appear to be short-range ordered but long-range disordered. This interesting phenomenon can be explained by the application of the Hohenberg-Mermin-Wagner theorem. In an ideal 2D material, the energy required for an increase in entropy (*i.e.*, promoting disorder) is relatively small, mainly because the energy associated with particle displacements is confined within the 2D plane. Consequently, the system is more prone to form the density fluctuations of thermodynamics [44–46]. As a result, strict long-range crystalline order cannot be achieved in a 2D system; instead, the system adopts a disordered or fluctuating state.

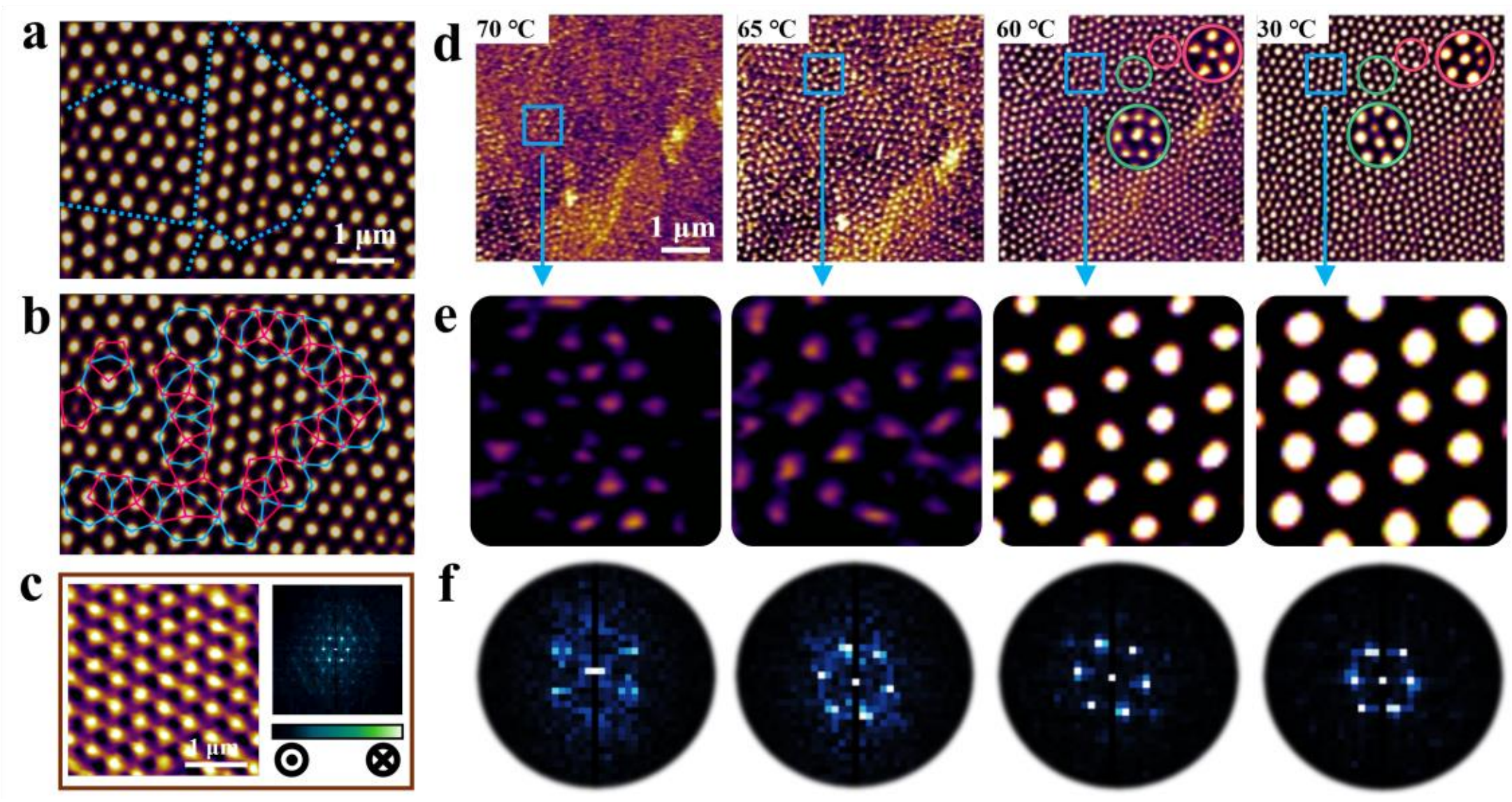

**Fig. 4.** Arrangement and rotational symmetry of skyrmions. (a) Skyrmion lattice composed of skyrmion clusters. The clusters are separated by blue dashed lines. (b) Skyrmion arrangement at boundaries. (c) The skyrmion arrangement within a cluster (left panel) and the corresponding Fast Fourier transform (right panel). (d) MFM images during the field cooling from 70 °C to 30 °C. The skyrmions in the red and green circles are enlarged. (e) The enlarged views of the skyrmion arrangement in the blue box within (d), and its corresponding Fast Fourier transform maps (f).

A region from MFM image of skyrmion lattice, obtained from FC to room temperature, was selected to study the arrangement of skyrmion, as shown in Fig. 4a-b (from one identical region). In Fig. 4c, a specific domain was chosen for magnification and subjected to a Fast Fourier transform (FFT) analysis. The examination reveals a sixfold coordination arrangement of skyrmions within this domain. The FFT plot prominently exhibits a sixfold rotational symmetry, further confirming a short-range ordering. However, the skyrmions between neighboring clusters exhibit different orientations. These distinct orientations give rise to dislocation-induced boundaries, represented by blue dashed lines in Fig. 4a. Skyrmions at the boundaries adopt a 5-fold or 7-fold coordinated configuration, represented by the blue pentagon and red heptagon in Fig. 4b, respectively. This arrangements are commonly referred to as 5/7 defects, which can be alternatively characterized as wedge disclination dipoles [47]. The formation of defects can be attributed to skyrmion clusters merging and relaxation.

Fig. 4d illustrates the generation of the skyrmion lattice during the FC process. At 70 °C, skyrmions begin to nucleate. Various clusters are formed at different nucleation centres. As the temperature decreases to 65 °C, clusters gradually grow and merge, accompanied by the generation of numerous skyrmions. However, the overall arrangement of skyrmions appears disordered. As the temperature is further decreased to 60 °C, neighboring clusters grow in contact and form dislocations. The densely arranged clusters constitute a skyrmion lattice. Upon cooling to room temperature (30 °C), the skyrmion lattice was further ordered by generated individual skyrmion. New skyrmions are generated at dislocations, which alleviates the misalignment of skyrmion arrangement at the boundaries. The 5-fold coordination defect serves as the preferred location for skyrmion generation through the monomer-by-monomer incorporation mechanism. As shown by the red circle and its enlargement in Fig. 4d, a skyrmion is generated at the 5-fold coordination defect as the temperature decreases from 60 °C to 30 °C, forming a stable arrangement of 6-fold coordinated skyrmions. Additionally, the skyrmion at the 7-fold coordination defect can split into two skyrmions with decreasing temperature, as demonstrated by the green circle and its enlarged view in Fig. 4d.

The region inside the blue box in Fig. 4d is magnified and presented in Fig. 4e, where the frequency range and offset are adjusted to eliminate background and emphasize skyrmions. It is evident that the skyrmion sizes gradually increase and exhibit a more orderly arrangement during the FC process. The corresponding FFT of MFM images in Fig. 4e are summarized in Fig. 4f. At 70 °C, skyrmions starts nucleating and dispersing sporadically, leading to irregular symmetry in the FFT. However, as the temperature decreases, the FFT images of the skyrmions exhibit ambiguous sixfold rotational symmetry at 65 °C, and a significant sixfold rotational symmetry at temperatures equal to or below 60 °C. The FFT results provide further evidence that the skyrmion arrangement gradually converges to an ordered sixfold coordination during the FC process. This ordering can be explained by Kosterlitz–Thouless–Halperin–Nelson–Young melting theory [48–52].

## 4. Conclusion

In summary, utilized both L-TEM and MFM real-space imaging techniques, we comprehensively investigate the generation of room-temperature skyrmions in $Fe_3GaTe_2$ via methods involving magnetic field induction, FC, and tip scanning. The DMI produced at the interface of oxide layers is the critical factor for the formation and stabilization Néel-type skyrmions. Through manipulating the magnetic field by utilizing magnetic scanning tip, a reversible transformation between labyrinthine domains and skyrmions can be readily achieved, endowing the remarkable capability for creating and erasure of skyrmion lattice. The skyrmion is gradually ordered during the FC process. The reversible creation and erasure of room-temperature skyrmion lattice in $Fe_3GaTe_2$ provides a solid foundation for developing advanced spintronic memory devices based on vdW magnetic materials.

## Acknowledgments

The work was supported by the National Natural Science Foundation of China (92163101, 12374080), the National Key Research and Development Program of China (2019YFA0307900), the Beijing Natural Science Foundation (Z190011) and the GL2022178037L. X.L.W acknowledges the National Natural Science Foundation of China (12074018). T.L.X. acknowledges the National Natural Science Foundation of China (12074425, 11874422) and the National Key Research and Development Program of China (2019YFA0308602). The authors would like to thank Shiyanjia Lab (www.shiyanjia.com) for the EDS measurement.

Supporting Information

# Local Manipulation of Skyrmion Lattice in $Fe_3GaTe_2$ at Room Temperature

Shuaizhao Jin [a], Zhan Wang [b], Shouzhe Dong [c], Yiting Wang [d,e], Kun Han [d,e], Guangcheng Wang [f], Zunyi Deng [a], Xingan Jiang [a,g], Ying Zhang [b], Houbing Huang [c], Jiawang Hong [a], Xiaolei Wang [f,*], Tianlong Xia[d,e,*], Sang-Wook Cheong [h], Xueyun Wang [a.*]

## 1. Crystal and Magnetic Measurements

The morphology, crystal structure, elemental distribution and ratios of $Fe_3GaTe_2$ were investigated using optical microscopy (OM, Axio Imager A2m), powder X-ray diffraction (XRD, Rigaku Ultima IV) with Cu Kα radiation (wavelength = 0.15406 nm) and Field Emission Scanning Electron Microscope (FE-SEM, TESCAN MIRA LMS) equipped with energy dispersive X-ray spectroscopy (EDS), respectively. These measurements were performed at room temperature. Magnetic measurements of $Fe_3GaTe_2$ crystals were carried out employing a Quantum Design magnetic property measurement system (QD MPMS-3).

## 2. Lorentz electron transmission microscopy

The magnetic domain structures were studied by using a JEOL-dedicated Lorentz TEM (JEOL 2100F). Double tilt heating holder (Gatan 652 TA) was used for high-temperature manipulation. The external perpendicular magnetic field was introduced by gradually increasing the objective lens current. The magnetic domain wall contrast at different focus was imaged under the convergent or divergent electron beam, which is introduced by the interaction of electron beam with the in-plane magnetization. The specimen along [001] zone axis for L-TEM observation was prepared via FIB milling.

## 3. Magnetic force microscopy measurement

The magnetic force microscopy measurements of variable temperature were performed by an atomic force microscope (Asylum Research MFP-3D Classic). A Si cantilever tip with Co-Cr coating was used, and the spring constant and resonance frequency were ~3 N $m^{-1}$ and ~75 kHz, respectively. The MFM mode is measured in tapping/lift mode. The temperature variation in the MFM test was achieved using Asylum Research's PolyHeater accessory.

## 4. Device fabrication and Hall measurement

The thin $Fe_3GaTe_2$ flakes were achieved through mechanical exfoliation from synthetic bulk crystals onto $SiO_2$ (300 nm)/Si substrates transferred by PDMS (polydimethylsiloxane). By using photolithography and magnetron sputtering, the Hall bar devices were fabricated with four Au electrodes on positioned $Fe_3GaTe_2$ flake, forming a 34 $\mu$m wide and 34 $\mu$m long channel with perpendicular voltage detection. Four-probe electrical measurements of the transverse/Hall resistance ($R_{xy}$) dependent on magnetic field were conducted in an electromagnet system, with a fixed 1 mA DC current applied on the channel.

## 5. Phase field simulation

In the phase-field mode, the spatial distribution of the magnetization field $\boldsymbol{M} = \boldsymbol{M}_s (m_1, m_2, m_3)$ is used to describe the magnetic domain structure, where $m_i$ represents the component of $\boldsymbol{m}$ along the local crystallographic coordinate axis i (i = $x, y, z$). $\boldsymbol{M}_s$ is the saturation magnetization and the components of unit magnetization vector, respectively. The transient magnetization domain structure is described by the Landau-Lifshitz-Gilbert (LLG) equation:

$$(1+\alpha^2)\frac{\partial \boldsymbol{M}}{\partial t} = -\gamma_0 \boldsymbol{M} \times \boldsymbol{H}_{\text{eff}} - \frac{\gamma_0 \alpha}{\text{M}_s} \boldsymbol{M} \times (\boldsymbol{M} \times \boldsymbol{H}_{\text{eff}}) \tag{1}$$

where $\alpha$ is the damping constant, and $\gamma_0$ is the gyromagnetic ratio. $H_{\text{eff}}$ is the effective magnetic field, which can be given as

$$H_{\text{eff}} = \frac{-1}{\mu_0 M_s} \frac{\partial F_{\text{tot}}}{\partial \boldsymbol{m}} \tag{2}$$

where $\mu_0$ and $F_{\text{tot}}$ denote vacuum permeability and total free energy, respectively. The total free energy can be written as

$$F_{\text{tot}} = F_{\text{dem}} + F_{\text{ani}} + F_{\text{exc}} + F_{\text{dmi}} \tag{3}$$

where $F_{\text{dem}}$, $F_{\text{ani}}$, $F_{\text{exc}}$, and $F_{\text{dmi}}$ are the demagnetization energy, magnetocrystalline anisotropy energy, exchange energy and DMI energy, respectively. The demagnetization energy of a system can be written as

$$F_{dem} = -\frac{1}{2}\mu_0 M_s \iiint H_{\text{dem}} \cdot \boldsymbol{m} \, dV \tag{4}$$

where $H_{\text{dem}}$ is the demagnetization field, determined by the long-range interaction among the magnetic moments in the system. The uniaxial anisotropy in ultra-thin films with a magnetic easy/hard axis perpendicular to the film plane. The volume of uniaxial anisotropy energy is given by

$$F_{\text{ani}} = \iiint [K_1(1-m_3^2) + K_2(1-m_3^2)^2] dV \tag{5}$$

where $K_1$ and $K_2$ are the uniaxial anisotropy constant. The exchange energy is determined by the spatial variation of the magnetization orientation and can be described as

$$F_{\text{exc}} = A\iiint ((\nabla m_1)^2 + (\nabla m_2)^2 + (\nabla m_3)^2) dV \tag{6}$$

where $A$ is the nearest-neighbor Heisenberg exchange coupling energy constants. The DMI energy arises from the interface with oxides, the interfacial DMI free energy is given as

$$F_{\text{dmi}} = \iiint D(m_3 \nabla \cdot \boldsymbol{m} - \boldsymbol{m} \cdot \nabla m_3) dV \tag{7}$$

where $D$ is the effective interfacial DMI constants.

The materials parameters of the vdW magnet $Fe_3GaTe_2$ used in this simulations are $A_{ex}$ = $4 \times 10^{-12}$ J $m^{-1}$, $K_1 = 6 \times 10^4$ J $m^{-3}$, $K_2 = 0$, $M_s = 2.3 \times 10^5$ A $m^{-1}$, $D = 0.4$ mJ $m^{-2}$; $\gamma_0 = 2.2 \times 10^5$ $mA^{-1}$ $s^{-1}$. The system is described by 3D discretized cells of 400△x × 400△y × 20△z, where a cell size (△x, △y, △z) = (5 nm, 5 nm, 5 nm) is used.

## 6. Magnetic anisotropy of the $Fe_3GaTe_2$ crystals

The magnetic anisotropy of bulk $Fe_3GaTe_2$ crystals is investigated through magnetic field-dependent magnetization measurements at 300 K. As depicted in Fig. 1f, $Fe_3GaTe_2$ bulk crystals exhibit distinct out-of-plane magnetic anisotropy, characterized by an in-plane saturation field of $B_{sat}$ = 4 T and a saturation magnetization of $M_{sat}$ = 28.3 emu/g (0.26 T). The corresponding magnetic anisotropy energy density ($K_u$) is determined to be approximately $4.15 \times 10^5$ J/$m^3$ at 300 K, calculated using the formula [1]:

$$K_u = \frac{B_{sat} M_{sat}}{2u_0}$$

where $M_{sat}$, $B_{sat}$ and $\mu_0$ are the saturation magnetization, the saturation field, and the permeability in free space, respectively.

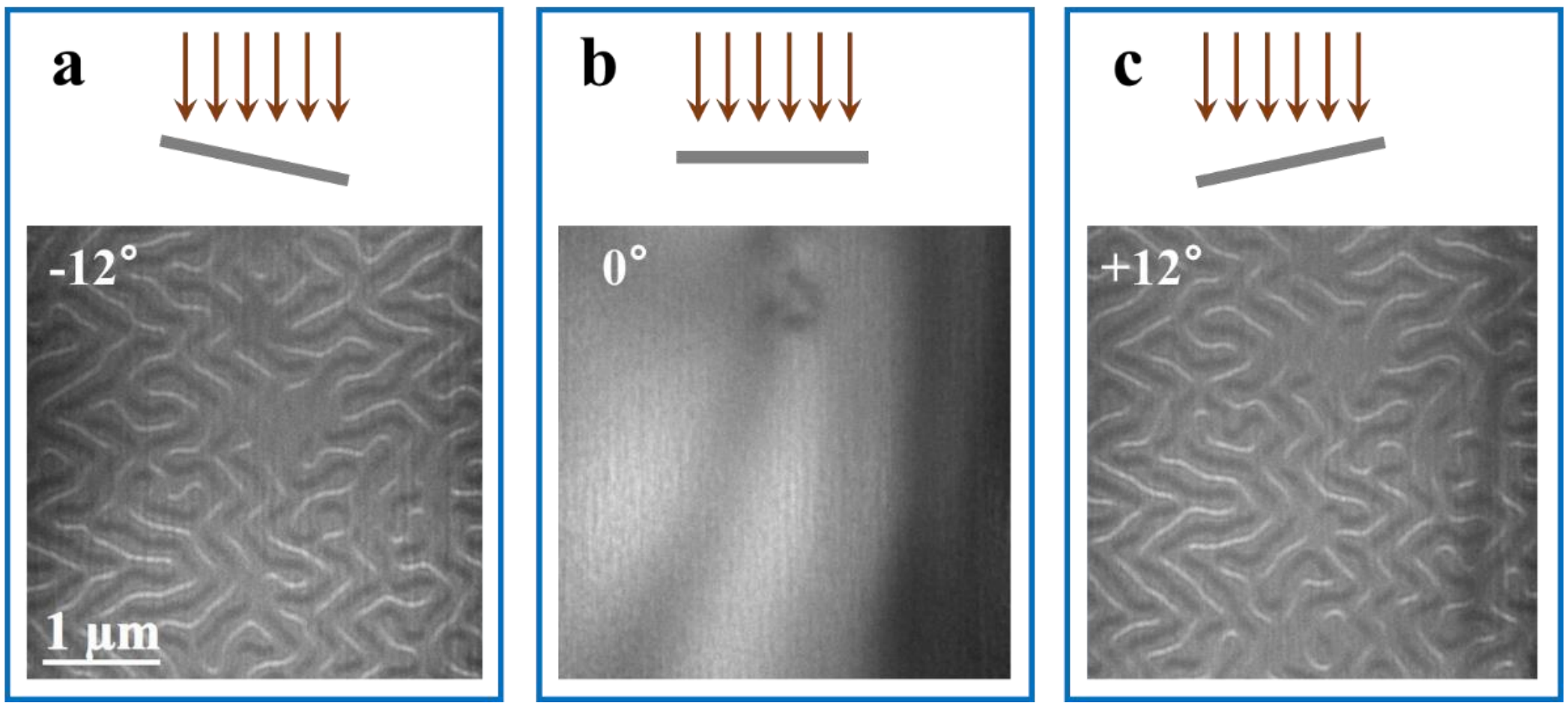

**Fig. S1**. **Labyrinthine domains images of sample with different tilt angles.** Labyrinthine domains of sample with tilted by (a) -12°, (b) 0° and (c) +12°. No magnetic domains are observed at a tilt angle of 0°, but labyrinthine domains become apparent at tilt angles of ±12°. This suggests the labyrinthine domain wall follows a Néel-type pattern.

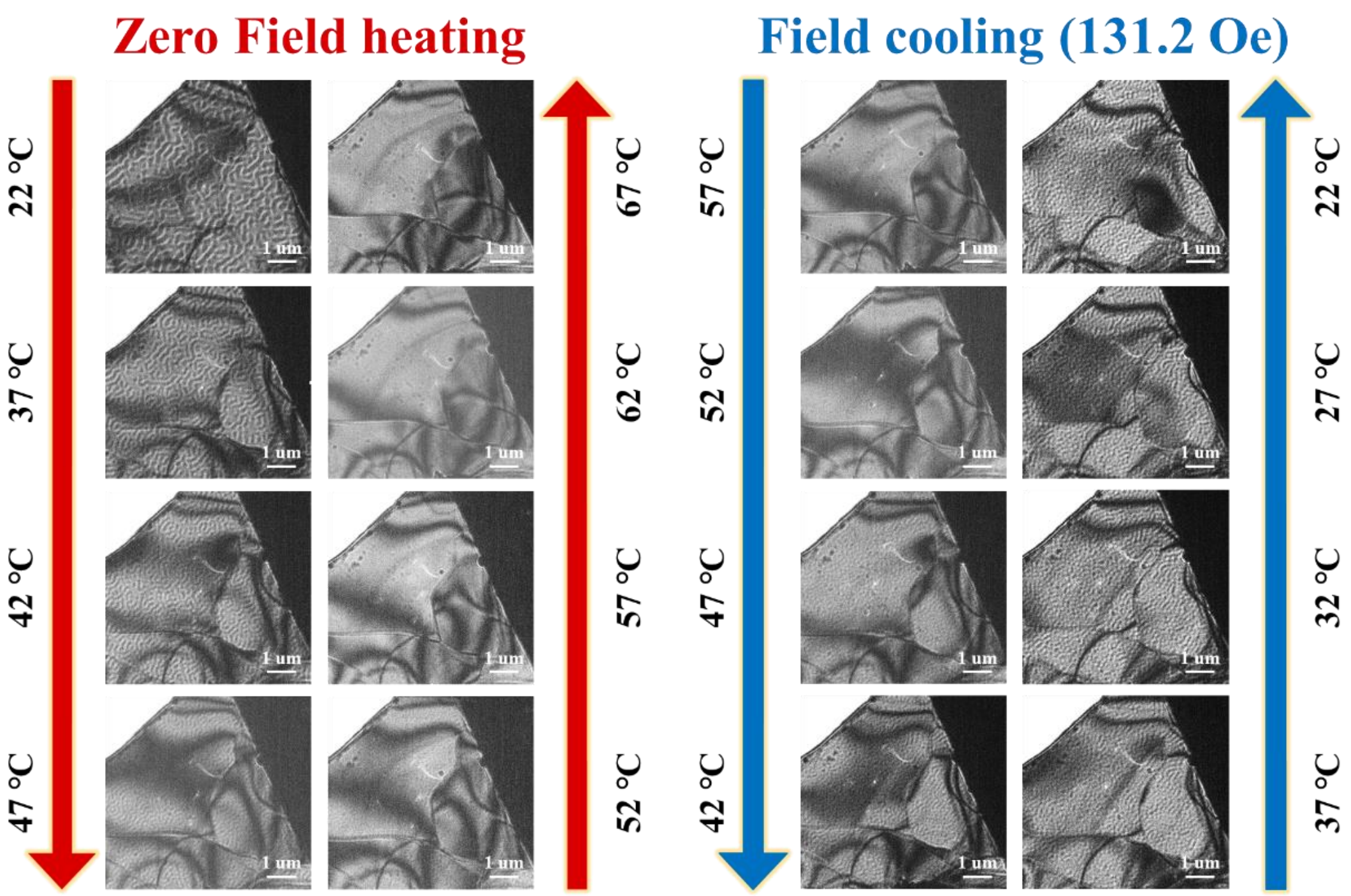

**Fig. S2. L-TEM images of the zero-field heating (left) and field-cooling (right) processes.** The applied magnetic field is 131.2 Oe. During zero-field heating, the signal of the labyrinthine domains gradually weakened and disappeared as the temperature increased. During field cooling, skyrmions gradually appear with decreasing temperature.

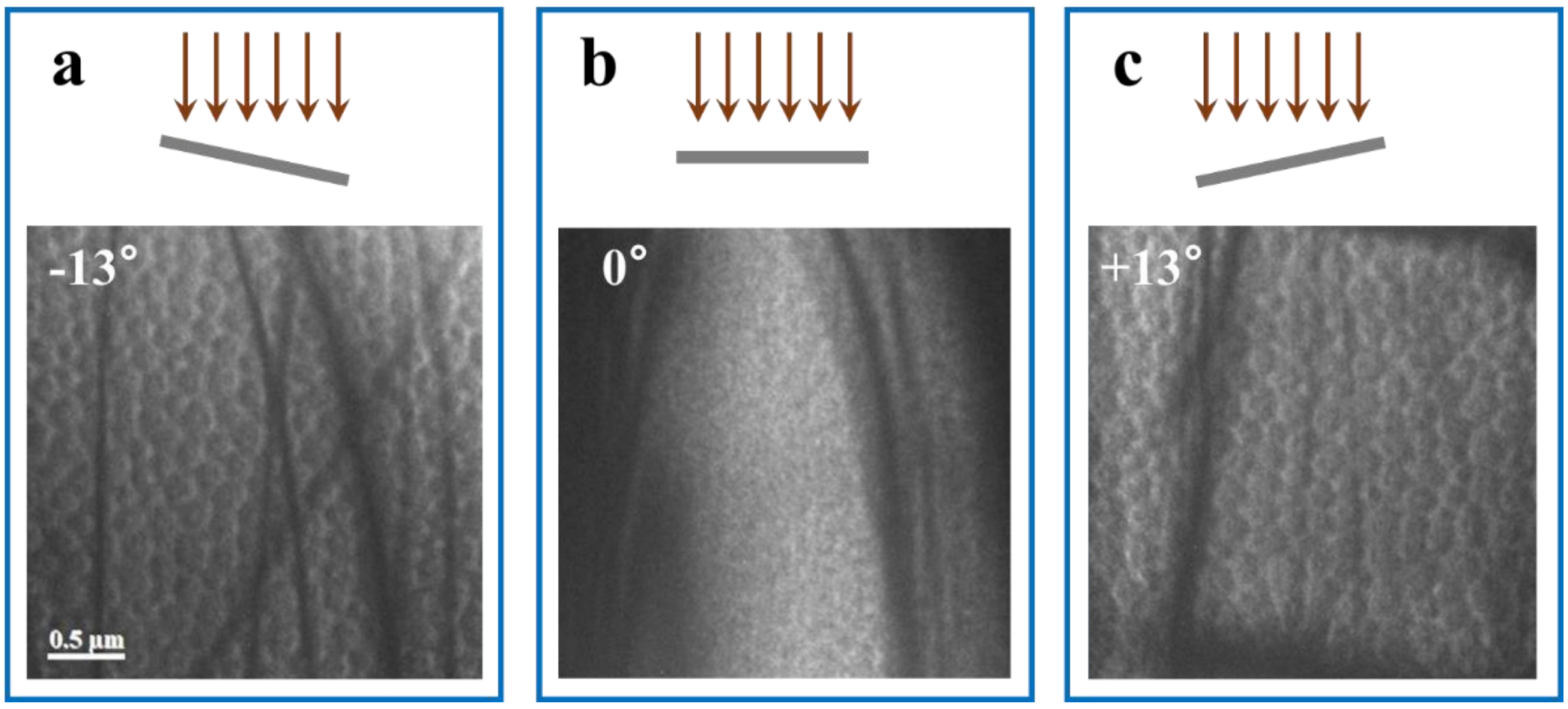

**Fig. S3**. **L-TEM images of Néel-type skyrmions with different sample tilt angles.** Néel-type skyrmions with sample tilted by (a) -12°, (b) 0° and (c) +12°. Skyrmions are absent at a tilt angle of 0°; however, they manifest distinctly at tilt angles of ±12°. This observation implies that the skyrmions conform to a Néel-type pattern.

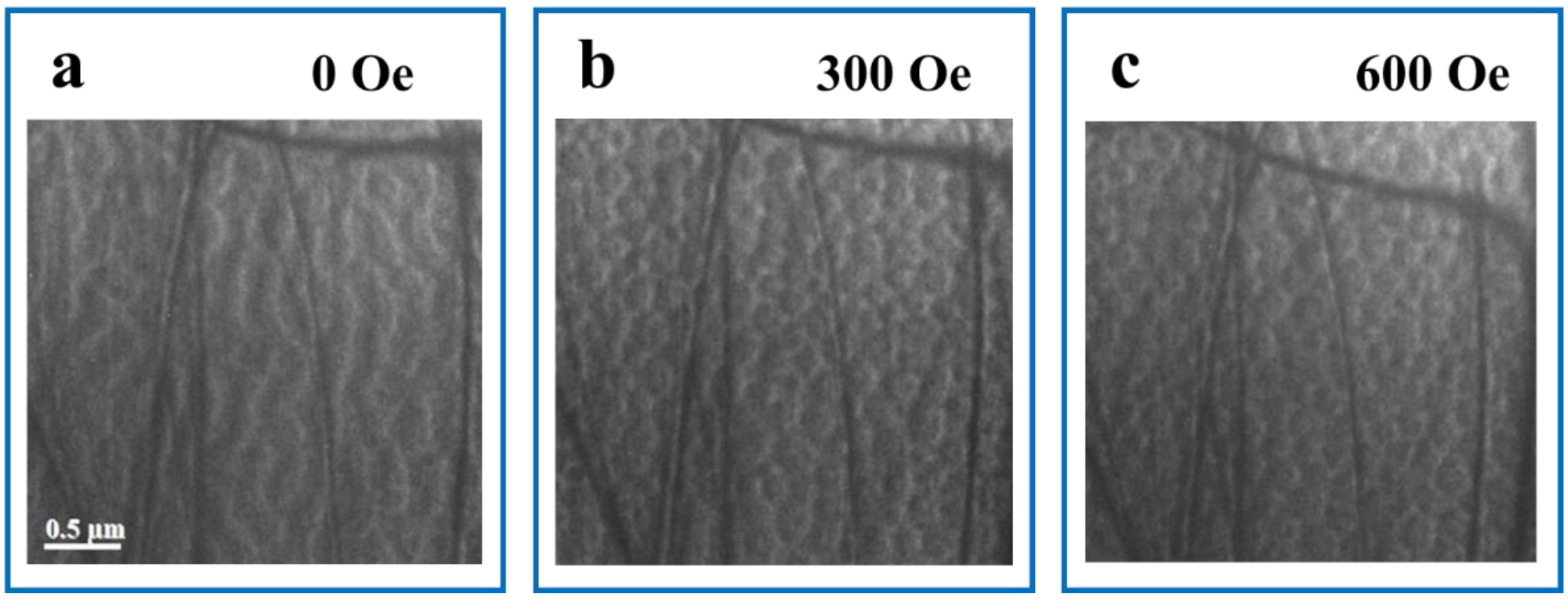

**Fig. S4**. **LTEM images obtained by field cooling.** Magnetic domains obtained by field cooling at (a) 0 Oe, (b) 300 Oe and (c) 600 Oe external magnetic fields, respectively. The density of the skyrmion lattice remained fairly consistent as the sample was cooled under 300 Oe and 600 Oe.

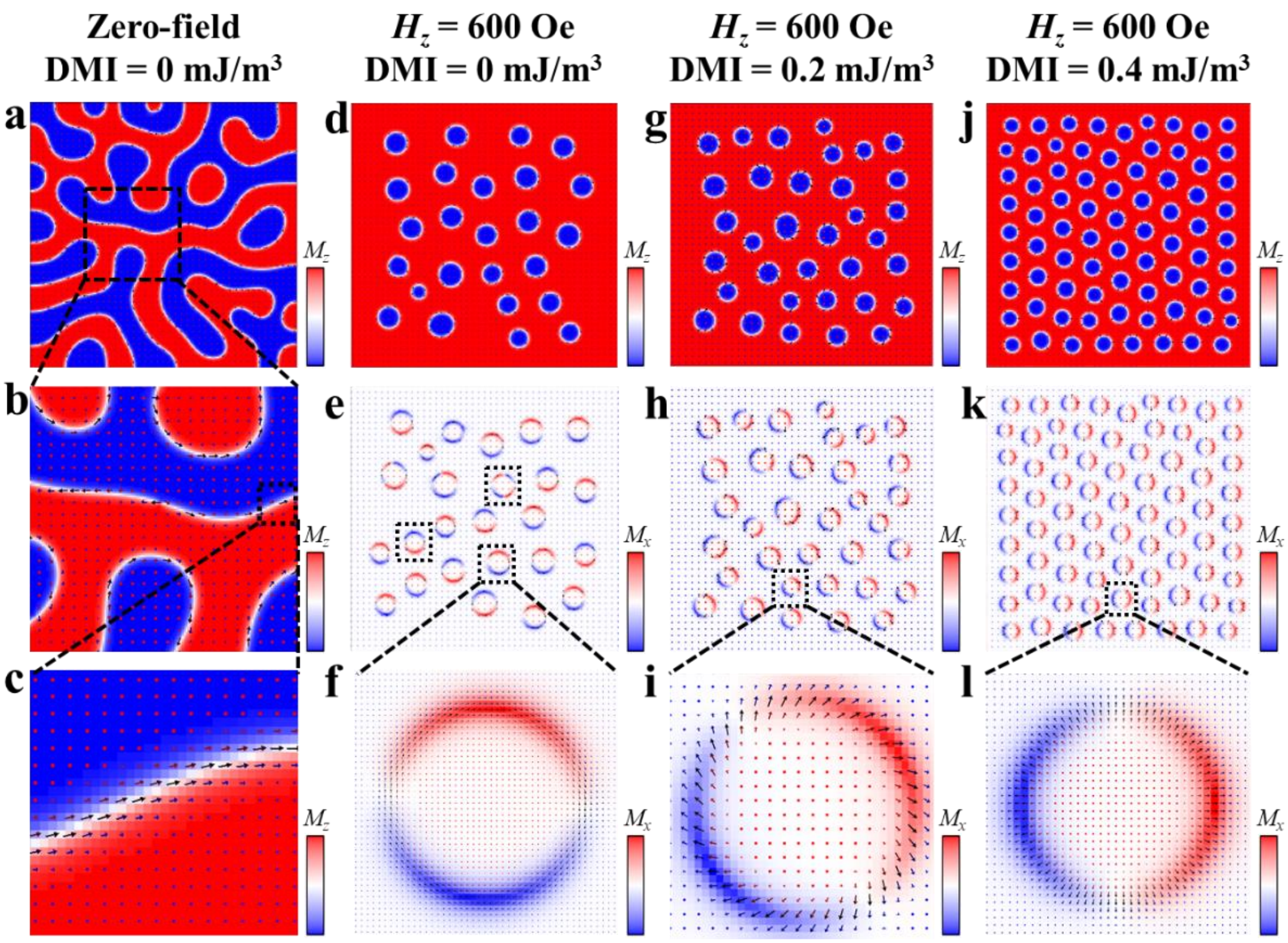

**Fig. S5**. **Simulated images under different conditions.** (a-c) Labyrinth domains and their enlarged views at 0 field and 0 DMI value. (d-f) Bloch skyrmions and their enlarged view at 600 Oe field and 0 DMI value. (g-i) Intermediate states and their enlarged view of skyrmion transition from Néel-tape to Bloch-tape at 600 Oe field and 0.2 mJ/m$^3$ DMI value. (j-l) Néel skyrmions and their enlarged view at 600 Oe field and 0.4 mJ/m$^3$ DMI value. The skyrmions created in $Fe_3GaTe_2$ gradually transform from the Bloch type to the Néel type by the introduction of DMI value.

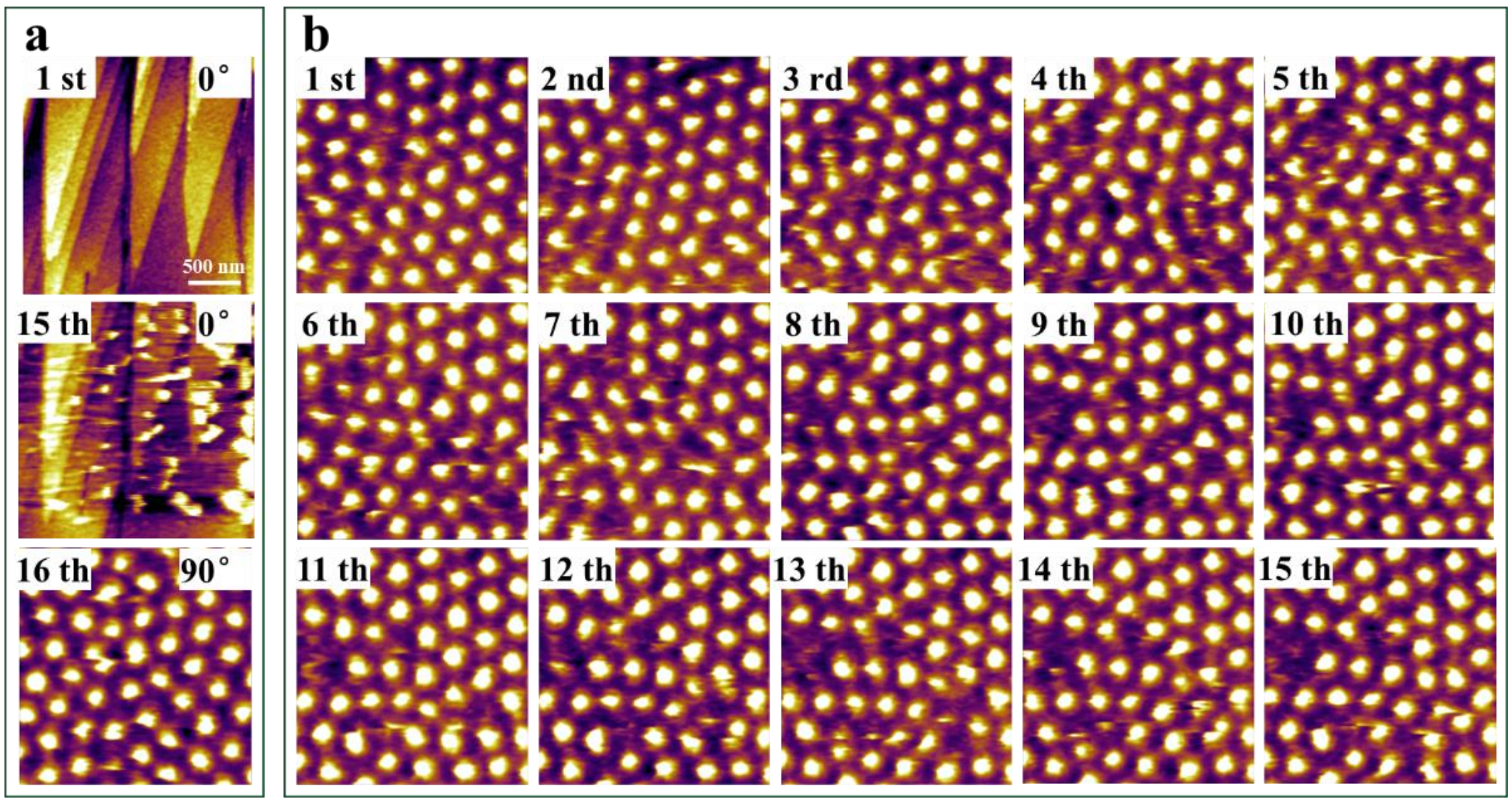

**Fig. S6. MFM images at different scan times and angles (the angle between tip scanning direction and cantilever) in the same region.** (a) The first scan of topography image (scanning angle 0°), the 15th scan of topography image (scanning angle 0°) and the 16th scan of magnetic domain image (scanning angle 90°). (b) MFM images were repeated 15 times at the scan angle of 0°. Note that the original state of $Fe_3GaTe_2$ shows zero-field skyrmion lattice at room temperature. The sample topography is destroyed but the skyrmion lattice remains almost unchanged after multiple scans. The skyrmion lattice remains stable after changing the scanning angle.